\documentclass[10pt,superscriptaddress,twocolumn]{revtex4-2}

\usepackage{tikz}
\usepackage[cmex10]{amsmath}
\usepackage{amssymb}
\usepackage{physics}
\usepackage{upgreek}
\usepackage{titlesec}
\usepackage{siunitx}
\usepackage{here}
\usepackage{tabularx}
\usepackage{color}
\usepackage{comment}
\usepackage{graphicx}

\usepackage{xcolor}
\usepackage{ulem}

\usepackage{etoolbox}
\makeatletter
\renewcommand\thetable{\arabic{table}}            
\renewcommand\tablename{Table}                   
\patchcmd{\@caption}
  {\tablename\ \thetable.}
  {\tablename~\thetable:}
  {}{}
\makeatother

\usepackage{hyperref}
\hypersetup{
    colorlinks=true,
    linktocpage=true,
    linkcolor=blue,
    filecolor=magenta,      
    urlcolor=cyan,
    bookmarks=true,}
    
\graphicspath{{figures/}}

\begin{document}

\title{Multilevel Effects and the Limits of the Quantum Rabi Model in Ultrastrong Coupling system}

\author{A. Tomonaga}\email{akiyoshi.tomonaga@aist.go.jp}
\affiliation{Institute of Advanced Industrial Science and Technology (AIST), Tsukuba, Ibaraki 305-8563, Japan}
\affiliation{Department of Physics, Tokyo University of Science, 1--3 Kagurazaka, Shinjuku, Tokyo 162--0825, Japan}
\affiliation{RIKEN Center for Quantum Computing (RQC), 2--1 Hirosawa, Wako, Saitama 351--0198, Japan}

\begin{abstract}

The spectrum of artificial atoms coupled to resonators in the ultrastrong coupling regime have been analyzed using the quantum Rabi model over a wide range of coupling strengths.
However, the accuracy of models obtained by truncating multilevel artificial atoms to two levels, as well as the correspondence between effective parameters extracted from spectral fitting and the bare parameters of the circuit Hamiltonian, depends on the multilevel structure of the system.
In this study, I analyze the spectrum of an ultrastrongly coupled circuit consisting of a multilevel superconducting artificial atom and a resonator using a circuit Hamiltonian, and investigate the effects of truncating higher atomic levels on the low-energy spectrum.
Based on L\"owdin partitioning, I evaluate the contributions of virtual transitions through higher atomic levels and show that diagonal energy corrections reduce the errors arising from simple two-level truncation over a wide parameter range.
I also show that there are parameter regimes in which readjustment of the effective Rabi parameters successfully reproduces the multilevel spectrum, while systematic deviations remain in other regimes.
These results demonstrate the necessity of separately evaluating two-level truncation, perturbative corrections, and effective-parameter fitting in ultrastrongly coupled systems, and provide guidance for quantitative spectral analysis of multilevel quantum circuits.

\end{abstract}

\maketitle

\section{Introduction}
\begin{figure*}[t]
\centering
\includegraphics[width=\textwidth]{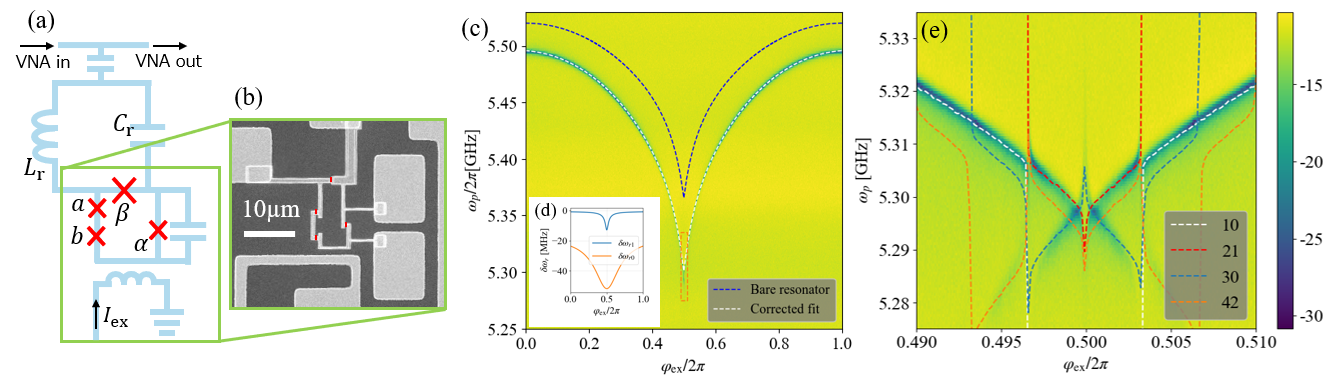}
\caption{
(a) Schematic of the measured circuit.
A capacitively shunted flux qubit (artificial atom) is galvanically coupled to an LC resonator through the $\beta$ junction.
The resonator is capacitively coupled to a transmission line for microwave measurements.
(b) Scanning electron microscope (SEM) image of the artificial atom.
The Josephson junctions are highlighted in red.
An on-chip flux-bias line controls the external magnetic flux applied to the artificial atom.
(c) Flux-dependent resonator spectrum measured through the microwave transmission coefficient $S_{21}$ using a vector network analyzer.
The white dotted line shows the fit obtained from Eq.~\eqref{resonator}, while the black dotted line shows the semiclassical result without the phenomenological frequency correction.
(d) The two phenomenological frequency-shift contributions $\delta\omega_0/2\pi$ and $\delta\omega_1/2\pi$ used in panel (c).
(e) High-resolution spectrum near half a flux quantum.
The dotted lines indicate transition frequencies calculated from the circuit Hamiltonian in Eq.~\eqref{HAR}.
The fitted circuit parameters are $E_{\mathrm J}/h=236$~GHz, $E_{\mathrm c}/h=2.92$~GHz, $\alpha=0.57$, $\beta=1.37$, $\gamma=3.57$, $L_{\mathrm r}=1.59$~nH, and $C_{\mathrm r}=414$~fF.
}
\label{fig1}
\end{figure*}

In superconducting quantum circuits, the energy structure of artificial atoms can be designed with considerable flexibility, allowing the realization of anharmonicities and coupling strengths that differ from those of systems based on natural atoms.
Owing to this high degree of design flexibility, various quantum phenomena have been reported in the ultrastrong coupling (USC) regime, including the verification of selection rules associated with parity symmetry \cite{wang2023probing}, the observation of spectra corresponding to the simultaneous excitation of two atoms by a single photon \cite{tomonaga_spectral_2025}, the realization of the deep-strong coupling regime \cite{yoshihara_superconducting_2017,forn-diaz_ultrastrong_2017}, and strong light--matter coupling using transmon artificial atoms with low anharmonicity \cite{bosman_multi-mode_2017,ye_near-ultrastrong_2025}.
To extend these achievements to quantum computation and quantum error correction utilizing ultrastrong coupling \cite{kyaw_creation_2015,Wang2016Holonomic,stassi_scalable_2020,chen_fast_2021}, the utilization of ground-state entanglement \cite{jaako_ultrastrong-coupling_2016,PhysRevLett.119.183602}, and quantum sensing \cite{cxvs-5pb1}, it is important to quantitatively understand the energy structure of these systems.

When the low-energy degrees of freedom of an artificial atom can be approximated by a two-level system, the coupled artificial atom--resonator system can be described over a wide range of coupling strengths by the following Rabi Hamiltonian \cite{niemczyk_circuit_2010,PhysRevLett.120.183601,yoshihara_characteristic_2017}:
\begin{equation}
\hat{\mathcal{H}}_{\mathrm{Rabi}}/\hbar
=
\frac{\varepsilon\sigma_z+\Delta\sigma_x}{2}
+\omega_{\mathrm r}\hat a^\dagger\hat a
+g\sigma_z(\hat a^\dagger+\hat a).
\label{rabi}
\end{equation}
Here, $\hbar\Delta$ is the minimum energy gap of the artificial atom, $\hbar\varepsilon$ is the energy bias controlled by the external magnetic flux, $\omega_{\mathrm r}$ is the angular frequency of the resonator, and $g$ is the coupling strength expressed in angular frequency units.
The transition angular frequency of the artificial atom is given by $\omega_{\mathrm q}=\sqrt{\varepsilon^2+\Delta^2}$.

However, even when the Rabi model accurately reproduces experimental spectra, its effective parameters do not necessarily coincide with the bare parameters defined by the multilevel circuit Hamiltonian.
Indeed, previous studies have reported that readjustment of the parameters in the Rabi Hamiltonian is required to reproduce the spectra obtained from circuit models \cite{tomonaga_quasiparticle_2021,yoshihara_hamiltonian_2022}.

Recently, Lamberto et al. proposed the Renormalized Quantum Rabi Model (RQRM), which perturbatively incorporates the contributions of higher atomic levels using the Schrieffer--Wolff transformation \cite{lamberto_renormalization_2025}.
In this effective model, renormalization of the atomic transition frequency and coupling strength is accompanied by additional terms, including terms quadratic in the resonator operators.
Numerical calculations for a one-dimensional double-well potential and a fluxonium--resonator system demonstrated that the RQRM improves the accuracy of the energy spectra compared with the conventional Rabi model.
Furthermore, the importance of extending this approach to flux qubits with multidimensional potentials has been pointed out.

On the other hand, for actual multijunction superconducting circuits, it is important to systematically investigate, based on experimental results, the extent to which higher atomic levels affect the spectra and the parameter regions in which two-level truncation and perturbative effective models remain quantitatively accurate.

In addition, the validity of the two-level approximation in the ultrastrong coupling regime has also been discussed from the perspective of gauge invariance \cite{noauthor_breakdown_nodate,di_stefano_resolution_2019}.
Even when Hamiltonians are related by unitary transformations and are equivalent in the complete Hilbert space, simply truncating the matter subsystem to two levels generally breaks the equivalence between gauges.
To address this issue, effective Hamiltonians that preserve consistency between gauges under the two-level approximation have been proposed.
However, gauge consistency and quantitative accuracy in reproducing the spectra of multilevel systems must be examined separately.
In particular, the contributions of virtual transitions through truncated higher atomic levels and the range over which these contributions can be incorporated perturbatively depend on the circuit representation and the parameter regime.

In this study, starting from circuit parameters determined from experimental spectra, I analyze the effects of truncating higher atomic levels on the low-energy spectrum using L\"owdin partitioning \cite{lowdin_note_1951}.
I compare a simple two-level model, diagonal energy corrections, and a second-order effective Hamiltonian including off-diagonal matrix elements.
I further investigate the accuracy obtained by refitting the effective parameters of the Rabi model.
By varying the circuit parameters and coupling strength, I systematically evaluate the errors of these approximations.
In addition, I analyze the resonator spectrum over a wide range of magnetic flux biases using a semiclassical model based on Josephson inductance combined with a phenomenological frequency correction.

\section{Circuit}

The measured circuit is shown in Fig.~\ref{fig1}.
An artificial atom is galvanically coupled to a lumped-element LC resonator through a Josephson junction.
The resonator is capacitively coupled to a transmission line, and the transmission coefficient $S_{21}$ was measured using a vector network analyzer (VNA).

To reduce the anharmonicity of the artificial atom, a capacitively shunted flux qubit is employed \cite{yan_flux_2016,PhysRevB.86.140508}.
The external magnetic flux is controlled by an on-chip flux-bias line.

I denote the Josephson energy of junction $a$ by $E_{\mathrm J}$ and assume that junction $b$ has the same size as junction $a$.
The potential energy of the artificial atom is expressed as
\begin{equation}
\begin{aligned}
\hat{\mathcal U}_{\mathrm J}
={}&-E_{\mathrm J}
\Bigl[
\beta\cos\hat\varphi_\beta
+\cos\hat\varphi_a\\
&\quad+\cos\bigl(
\varphi_{\mathrm{ex}}-\hat\varphi_a-\hat\varphi_\alpha-\hat\varphi_\beta
\bigr)
+\alpha\cos\hat\varphi_\alpha
\Bigr]
+E_{\mathrm L}\hat\varphi_\beta^2,
\end{aligned}
\label{eq:uj}
\end{equation}
where $\hat\varphi_m$ ($m\in\{a,\alpha,\beta\}$) represents the phase difference operator across each junction, and $\alpha$ and $\beta$ are the ratios of the Josephson energies of the corresponding junctions to that of junction $a$.
Here, $E_{\mathrm L}=(\Phi_0/2\pi)^2/(2L_{\mathrm r})$, where $\Phi_0$ is the magnetic flux quantum.

Using the charging energy $E_{\mathrm c}$ of junction $a$, the Hamiltonian of the artificial atom is given by
\begin{equation}
\hat{\mathcal H}_{\mathrm A}
=
4E_{\mathrm c}\hat{\boldsymbol n}^{\mathsf T}
\boldsymbol M^{-1}\hat{\boldsymbol n}
+\hat{\mathcal U}_{\mathrm J},
\label{eq:HA}
\end{equation}
where $\boldsymbol M$ is the normalized capacitance matrix (mass matrix),
\begin{align}
\boldsymbol M
=
\begin{pmatrix}
2 & 1 & 1\\
1 & \gamma+1 & 1\\
1 & 1 & \beta+1
\end{pmatrix},
\label{eq:matrix}
\end{align}
and $\hat{\boldsymbol n}$ is the vector of Cooper-pair number operators conjugate to the phase variables \cite{Tomonaga_2025}.
The parameter $\gamma$ is a normalized parameter associated with the shunt capacitance.
In the present circuit, $\gamma=\alpha$ corresponds to the absence of an additional shunt capacitor.

The eigenstates $\ket{\psi_i}$ and eigenangular frequencies $\omega_i$ of the artificial atom are obtained numerically from
$\hat{\mathcal H}_{\mathrm A}\ket{\psi_i}=\hbar\omega_i\ket{\psi_i}$.
The Hamiltonian of the coupled artificial atom--resonator system is then given by
\begin{equation}
\hat{\mathcal H}_{\mathrm{AR}}/\hbar
=
\sum_i\omega_i\ket{\psi_i}\bra{\psi_i}
-\sum_{i,j}g_{ij}\ket{\psi_i}\bra{\psi_j}
(\hat a^\dagger+\hat a)
+\omega_{\mathrm r}\hat a^\dagger\hat a,
\label{HAR}
\end{equation}
where the coupling matrix elements are expressed as
$\hbar g_{ij}
=
I_{\mathrm{zpf}}(\Phi_0/2\pi)
\mel{\psi_i}{\hat\varphi_\beta}{\psi_j}$ with $I_{\mathrm{zpf}}=
\sqrt{\hbar\omega_{\mathrm r}2L_{\mathrm r}}
$~\cite{Tomonaga_2025}.
Here, $I_{\mathrm{zpf}}$ is the zero-point fluctuation current of the resonator.
In the following, the transition angular frequency from the ground state to the $j$th excited state of the isolated artificial atom is denoted by
$\omega_{0j}^{(\mathrm A)}=\omega_j-\omega_0$.
I also define the normalized external magnetic flux as $f=\varphi_{\mathrm{ex}}/(2\pi)$.

Fig.~\ref{fig1}(c) shows the resonator spectrum measured over one period of the magnetic flux bias.
Except in the immediate vicinity of $f=0.5$, the observed response predominantly corresponds to the resonator mode.
Fig.~\ref{fig1}(e) shows the high-resolution spectrum near the flux bias where the transition frequency between the ground and first excited states of the artificial atom matches the resonator frequency.
The fitting results obtained using Eq.~\eqref{HAR} are indicated by dotted lines.
The obtained circuit parameters are
$E_{\mathrm J}/h=236$~GHz,
$E_{\mathrm c}/h=2.92$~GHz,
$\alpha=0.57$,
$\beta=1.37$,
$\gamma=3.57$,
$L_{\mathrm r}=1.59$~nH,
and $C_{\mathrm r}=414$~fF.

\begin{figure*}[t]
\centering
\includegraphics[width=\textwidth]{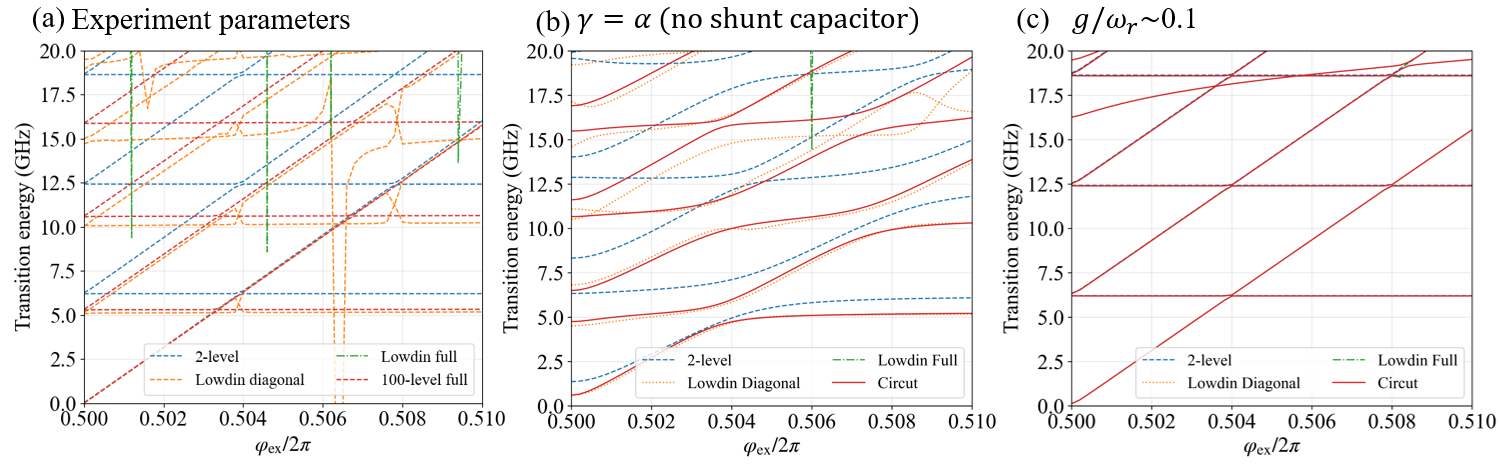}
\caption{
Comparison of transition spectra calculated using the full multilevel circuit Hamiltonian,
the two-level truncated model, and the second-order L\"owdin approximations.
``L\"owdin diagonal'' denotes state-specific diagonal energy corrections evaluated using
the full $Q$-space resolvent [Eq.~\eqref{eq:lowdin_second_order}],
whereas ``L\"owdin full'' denotes rediagonalization of the symmetrized
second-order effective Hamiltonian including off-diagonal matrix elements
[Eq.~\eqref{eq:lowdin_full}].
(a) Circuit parameters fitted to the experimental spectrum in Fig.~\ref{fig1}(e).
(b) Same as (a), but with $\gamma=\alpha$, corresponding to the absence of the additional shunt capacitance.
(c) Same as (a), but with all atom--resonator coupling matrix elements $g_{ij}$ reduced by a factor of eight.
}
\label{fig2}
\end{figure*}
\section{Two-level truncation}

In this section, I evaluate the spectral errors arising from truncating the artificial atom to two levels within the same circuit representation and investigate the contributions of virtual transitions through higher atomic levels.
Here, I distinguish between two-level truncation with fixed circuit parameters and refitting of the effective Rabi parameters.

\subsection{L\"owdin partitioning and second-order corrections}

Let $\hat H\equiv\hat{\mathcal H}_{\mathrm{AR}}$, I define the projection operators that retain the lowest two levels of the artificial atom as $\hat P=
\left(
\ket{\psi_0}\bra{\psi_0}
+\ket{\psi_1}\bra{\psi_1}
\right)\otimes\hat I_{\mathrm r}$
and
$\hat Q=\hat I-\hat P$,
where $\hat I_{\mathrm r}$ is the identity operator in the resonator Hilbert space.
The Hamiltonian can be written in block form as
\begin{equation}
\hat H=
\begin{pmatrix}
\hat H_{PP}&\hat H_{PQ}\\
\hat H_{QP}&\hat H_{QQ}
\end{pmatrix},
\label{eq:blockH}
\end{equation}
where $\hat H_{PP}=\hat P\hat H\hat P$,
$\hat H_{PQ}=\hat P\hat H\hat Q$,
$\hat H_{QP}=\hat H_{PQ}^\dagger$,
and $\hat H_{QQ}=\hat Q\hat H\hat Q$.
From Eq.~\eqref{HAR},
\begin{equation}
\hat H_{PQ}
=
-\hbar\sum_{i=0}^{1}\sum_{j\geq2}
g_{ij}\ket{\psi_i}\bra{\psi_j}
\otimes(\hat a^\dagger+\hat a).
\label{eq:HPQ}
\end{equation}

By eliminating the $Q$ space through L\"owdin partitioning, I obtain the energy-dependent effective Hamiltonian
\begin{equation}
\hat H_{\mathrm{eff}}(E)
=
\hat H_{PP}
+\hat H_{PQ}(E-\hat H_{QQ})^{-1}\hat H_{QP},
\label{eq:Heff_exact}
\end{equation}
\cite{lowdin_note_1951,feshbach_unified_1958,feshbach_unified_1962,lowdin_studies_1962}.
When $(E-\hat H_{QQ})^{-1}$ exists, solving
$\hat H_{\mathrm{eff}}(E)\ket{\chi}=E\ket{\chi}$
self-consistently with respect to $E$ is equivalent to solving the original eigenvalue problem for eigenstates with a nonzero projection onto the $P$ space.

In the following, I discuss the system using second-order approximations.
First, I diagonalize the Hamiltonian in the $P$ space as
$\hat H_{PP}\ket{u_n}=\epsilon_n\ket{u_n}$.
I define the matrix elements of the self-energy, including the internal interactions within the $Q$ space, as
\begin{equation}
\Sigma_{nm}(E)
=
\mel{u_n}{
\hat H_{PQ}(E-\hat H_{QQ})^{-1}\hat H_{QP}
}{u_m}.
\label{eq:Sigma}
\end{equation}
The diagonal correction evaluated at the unperturbed energy $\epsilon_n$ is given by
\begin{equation}
\Delta\epsilon_n^{(2)}
=
\Sigma_{nn}(\epsilon_n)
=
\mel{u_n}{
\hat H_{PQ}(\epsilon_n-\hat H_{QQ})^{-1}\hat H_{QP}
}{u_n}.
\label{eq:lowdin_second_order}
\end{equation}
Here, "second order" refers to the order in the coupling $\hat H_{PQ}+\hat H_{QP}$ between the $P$ and $Q$ spaces, with $\hat H_{PP}$ and $\hat H_{QQ}$ held fixed.

The Hermitian second-order effective Hamiltonian including off-diagonal corrections is constructed as
\begin{equation}
\bigl(H_{\mathrm{eff}}^{(2)}\bigr)_{nm}
=
\epsilon_n\delta_{nm}
+\frac{1}{2}
\left[
\Sigma_{nm}(\epsilon_n)
+\Sigma_{nm}(\epsilon_m)
\right]
\label{eq:lowdin_full}
\end{equation}
\cite{lowdin_studies_1962,bravyi_schriefferwolff_2011}.
Using
$\hat H_{QQ}\ket{v_k}=E_k^{(Q)}\ket{v_k}$
and
$W_{nk}=\mel{u_n}{\hat H_{PQ}}{v_k}$,
the correction term is expressed as
\begin{equation}
\Delta H_{nm}^{(2)}
=
\frac{1}{2}\sum_k W_{nk}W_{mk}^{*}
\left[
\frac{1}{\epsilon_n-E_k^{(Q)}}
+\frac{1}{\epsilon_m-E_k^{(Q)}}
\right].
\label{eq:lowdin_full_elements}
\end{equation}
For $n=m$, this expression reduces to Eq.~\eqref{eq:lowdin_second_order}.

Equation~\eqref{eq:Heff_exact} has the same projection form as the resolvent method used by Lamberto et al. to analyze higher-order corrections \cite{lamberto_renormalization_2025}.
Both the Schrieffer--Wolff transformation and L\"owdin partitioning aim to eliminate the degrees of freedom associated with higher atomic levels and incorporate their contributions into a low-energy effective Hamiltonian.
However, the specific approximation methods differ.
In the previous study, an effective Hamiltonian containing renormalized parameters and additional operator terms was constructed analytically through a perturbative expansion in the coupling to higher atomic levels.
In contrast, in the present study, I divide the circuit Hamiltonian into a low-energy subspace $P$ and a higher-energy subspace $Q$, and numerically evaluate the resolvent of the full $\hat H_{QQ}$, including the internal interactions within the $Q$ space.
This allows me to directly compare the accuracy of diagonal energy corrections for individual eigenstates with that of a second-order effective Hamiltonian including off-diagonal matrix elements.

In this paper, I refer to the method of calculating
$\epsilon_n+\Delta\epsilon_n^{(2)}$
based on Eq.~\eqref{eq:lowdin_second_order} as
"L\"owdin diagonal," and the method of rediagonalizing the second-order effective Hamiltonian constructed from Eq.~\eqref{eq:lowdin_full} as
"L\"owdin full."

When the diagonal energy corrections
$\Delta\epsilon_n^{(2)}$ are used, the transition angular frequency for $n>m$ is given by
\begin{equation}
\widetilde{\omega}_{nm}
=
\omega_{nm}^{(P)}
+\frac{\Delta\epsilon_n^{(2)}-\Delta\epsilon_m^{(2)}}{\hbar},
\end{equation}
where
\begin{equation}
\omega_{nm}^{(P)}=\frac{\epsilon_n-\epsilon_m}{\hbar}.
\label{eq:renormalized_transition}
\end{equation}
When the energy $\epsilon_n$ under consideration lies below all eigenvalues of the $Q$-space Hamiltonian, $\Delta\epsilon_n^{(2)}\leq0$.
This means that virtual transitions through higher atomic levels lower the energy of the corresponding state.
Furthermore, independently of perturbation theory, the eigenvalues obtained by projecting a given Hamiltonian onto a subspace provide upper bounds to the corresponding eigenvalues of the full Hamiltonian when both sets of eigenvalues are arranged in ascending order.
However, because the correction to a transition frequency is determined by the difference between two energy corrections, its sign is not generally fixed.
In Fig.~\ref{fig2}, simple two-level truncation tends to overestimate the transition frequencies.
This indicates that the excited states under comparison experience larger downward energy shifts than the reference state.


\subsection{Comparison of transition spectra}

Figure~\ref{fig2} shows a comparison of the transition spectra calculated using the approximate models and the full multilevel circuit Hamiltonian.
The reference spectra were obtained by numerical diagonalization of Eq.~\eqref{HAR}.
In the calculations, the Cooper-pair number for each junction was taken from $-8$ to $8$, and 100 eigenstates of the artificial atom and 25 Fock states of the resonator were retained \cite{Tomonaga_2025}.
The 2-level model represents the result obtained by projecting the artificial atom onto its lowest two levels without changing the circuit parameters.
I compare this result with the L\"owdin diagonal and L\"owdin full approximations defined in the previous subsection.

Figure~\ref{fig2}(a) shows the results obtained using the fitted parameters of the measured circuit.
Simple two-level truncation produces large spectral deviations from the full multilevel circuit Hamiltonian.
The L\"owdin diagonal approximation improves the agreement in the low-energy region, whereas substantial deviations remain at higher energies.
Even with the L\"owdin full approximation, the agreement cannot be uniformly improved over the entire spectrum considered.
Here, the coupling strength is $g_{01}/2\pi=5.45$~GHz, and the atomic transition frequency is $\omega_{02}^{(\mathrm A)}/2\pi=16.22$~GHz at $f=0.5$.

Figure~\ref{fig2}(b) corresponds to the case where the additional shunt capacitor is removed by setting $\gamma=\alpha$.
At $f=0.5$, the atomic transition frequency is $\omega_{02}^{(\mathrm A)}/2\pi=30.26$~GHz, and the coupling strength is $g_{01}/2\pi=4.81$~GHz.
Although the energy separation between the lowest two atomic levels and the higher levels is large, substantial spectral errors still appear in the L\"owdin full approximation.
This indicates that the accuracy of the second-order approximation employed here cannot be determined solely from the bare atomic level separations.

Figure~\ref{fig2}(c) shows the results obtained using the same atomic parameters as in Fig.~\ref{fig2}(a), but with all coupling matrix elements reduced according to $g_{ij}\rightarrow g_{ij}/8$.
In this case, $g_{01}/2\pi=0.68$~GHz at $f=0.5$, and the system remains in the ultrastrong coupling regime.
The atomic level structure is unchanged from that in panel (a), with $\omega_{02}^{(\mathrm A)}/(2\pi)=16.22$~GHz at $f=0.5$.
Reducing the coupling strength also decreases the errors arising from simple two-level truncation, and the spectra obtained with the L\"owdin corrections agree well with those of the full multilevel circuit Hamiltonian.
Therefore, the contributions of higher atomic levels are small in this regime and can be incorporated perturbatively.

\subsection{Fitting effective Rabi parameters}

\begin{figure}[t]
\centering
\includegraphics[width=85mm]{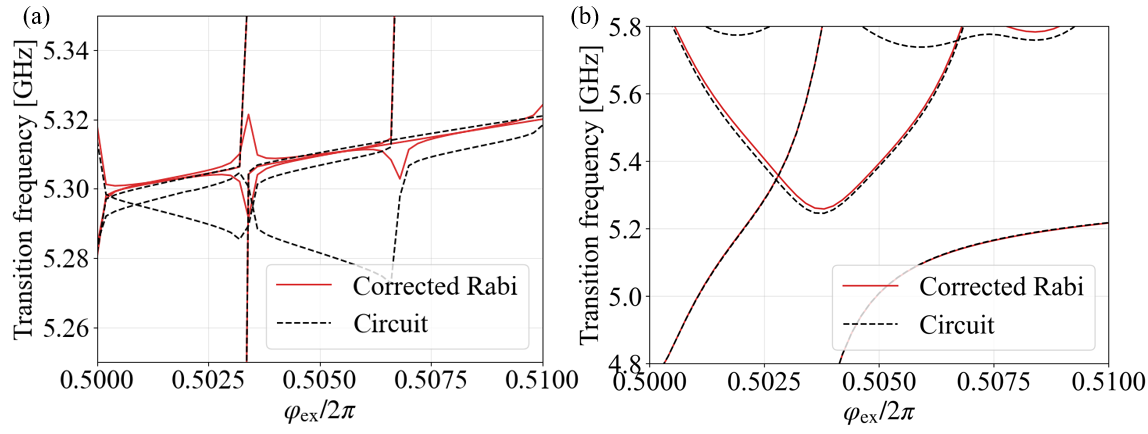}
\caption{
Comparison of transition spectra calculated using the full multilevel circuit Hamiltonian
and the Rabi model with fitted effective parameters.
The fitted Rabi model also includes the phenomenological flux dependence of the resonator frequency.
(a) Parameters corresponding to the measured circuit.
(b) Modified circuit parameters with $\gamma=\alpha$.
}
\label{fig3}
\end{figure}

Figure~\ref{fig3} shows the results of fitting the spectra of the full multilevel circuit Hamiltonian with the Rabi Hamiltonian in two parameter regimes.
Unlike simple two-level truncation, the effective parameters of the Rabi model are readjusted in this analysis.
Furthermore, I introduce a phenomenological correction describing the magnetic-flux dependence of the resonator frequency $\omega_{\mathrm r}'(f)=\omega_{\mathrm r}+A\varepsilon(f)$ following Ref.~\cite{tomonaga_quasiparticle_2021}.

For the measured circuit parameters shown in Fig.~\ref{fig3}(a), systematic deviations remain between the Rabi model and the full circuit Hamiltonian for the parameters and fitting range adopted in this study.
The parameters obtained from the fitting are
$\varepsilon/[2\pi(f-0.5)]=1.58\times10^3~\mathrm{GHz},$
$\Delta/2\pi=0.22~\mathrm{GHz}$,
$\omega_{\mathrm r}/2\pi=5.30~\mathrm{GHz}$,
$g/2\pi=6.80~\mathrm{GHz}$, and
$A=2.48\times10^{-4}$.
Here, $\varepsilon/[2\pi(f-0.5)]$ represents the linear slope of $\varepsilon/2\pi$ with respect to the magnetic flux bias.

On the other hand, Fig.~\ref{fig3}(b) shows the case of $\gamma=\alpha$.
Under this condition, the higher atomic levels are relatively well separated from the lowest two levels.
Although some residual deviations remain, the modified Rabi model reproduces the spectrum within the displayed range with good accuracy.
The obtained parameters are
$\varepsilon/[2\pi(f-0.5)]=1.44\times10^3~\mathrm{GHz}$,
$\Delta/2\pi=5.01~\mathrm{GHz}$,
$\omega_{\mathrm r}/2\pi=5.23~\mathrm{GHz}$,
$g/2\pi=5.41~\mathrm{GHz}$, and
$A=5.52\times10^{-4}$
In both cases, the effective resonator frequency of the Rabi model is lower than the bare resonator frequency $\omega_{\mathrm r}/2\pi=6.21$~GHz used in the simple two-level truncation.
Furthermore, the effective coupling strength $g/2\pi=6.80$~GHz in Fig.~\ref{fig3}(a) is larger than $g_{01}/2\pi=5.45$~GHz obtained from the circuit Hamiltonian.
However, the Rabi-model parameter $g$ and the coupling matrix element $g_{01}$ between specific atomic eigenstates are defined differently.

The bare parameters of the circuit Hamiltonian are defined within a representation that specifies how the system is partitioned into the artificial atom and the resonator.
In contrast, the fitted parameters of the Rabi model are effective quantities introduced to reproduce the spectrum of a multilevel system using a reduced set of degrees of freedom, and they do not necessarily coincide with the bare parameters.
When the representation and projection are fixed, the corrections arising from the elimination of higher atomic levels can be clearly defined.
However, parameter differences obtained from different representations or fitting conditions cannot be regarded as unique, representation-independent renormalization quantities.

\begin{figure*}[t]
\centering
\includegraphics[width=\textwidth]{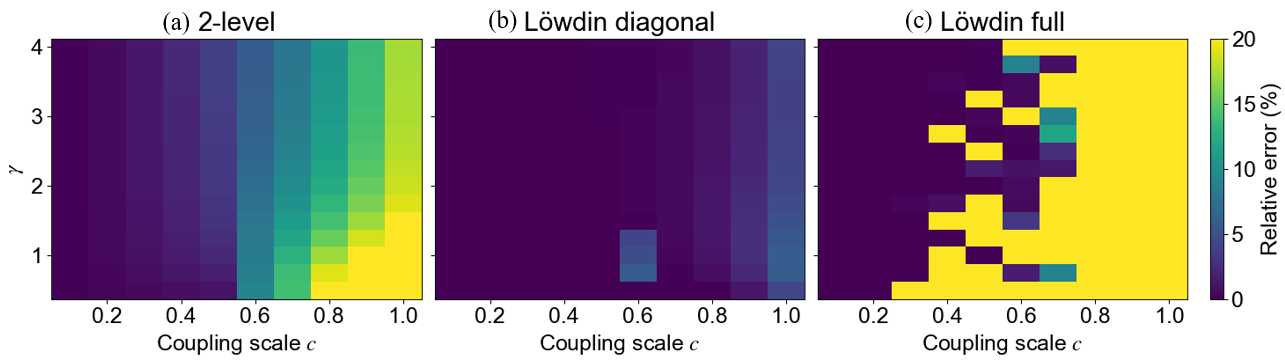}
\caption{
Relative errors in the effective resonator frequency, identified with the
$|0\rangle\rightarrow|2\rangle$ transition frequency $\omega_{02}$ throughout
the parameter range considered, with respect to the full multilevel circuit Hamiltonian.
(a) Two-level truncated model.
(b) State-specific diagonal energy corrections using the full $Q$-space resolvent
(L\"owdin diagonal).
(c) Rediagonalization of the symmetrized second-order effective Hamiltonian,
including off-diagonal matrix elements (L\"owdin full).
The horizontal axis is the coupling scale $c$, which uniformly changes all
atom--resonator coupling matrix elements as $g_{ij}\rightarrow cg_{ij}$.
The vertical axis is the circuit parameter $\gamma$.
Relative errors are defined in Eq.~\eqref{eq:relative_error}, and
values of 20\% or more are saturated at the upper limit of the common color scale.
}
\label{fig4}
\end{figure*}
\subsection{Parameter dependence of the resonator-frequency error}

Figure~\ref{fig4} shows the relative errors in the effective resonator frequency calculated using each approximate model as functions of the circuit parameter $\gamma$ and the coupling scale $c$.
The parameter $c$ is a numerical scaling factor applied to all atom--resonator coupling matrix elements calculated for each $\gamma$, according to $g_{ij}\rightarrow cg_{ij}$.
Here, $c=1$ corresponds to the unscaled coupling at each value of $\gamma$, and the parameters of the measured circuit correspond to $(c,\gamma)=(1,3.57)$.

Within the parameter range considered in this study, the transition corresponding to the resonator mode can always be identified as the transition $\omega_{02}$ from the ground state to the second excited state of the coupled system.
I therefore define the relative error in the effective resonator frequency $\omega_{02}^{(M)}$ obtained from model $M$ as
\begin{equation}
\eta_M
=100\left|
\frac{\omega_{02}^{(M)}-\omega_{02}^{(\mathrm{full})}}
{\omega_{02}^{(\mathrm{full})}}
\right|\quad[\%],
\label{eq:relative_error}
\end{equation}
where $M$ denotes one of the 2-level, L\"owdin diagonal, or L\"owdin full models, and $\omega_{02}^{(\mathrm{full})}$ is the corresponding transition angular frequency obtained from the full multilevel circuit Hamiltonian.
In Fig.~\ref{fig4}, values of $\eta_M\geq20\%$ are displayed using the same upper-limit color.

In the simple two-level truncation shown in Fig.~\ref{fig4}(a), the relative error in the resonator mode generally increases with increasing coupling scale $c$.
In contrast, the L\"owdin diagonal approximation in Fig.~\ref{fig4}(b) substantially reduces the error over a wide parameter range.
This result indicates that the energy corrections associated with virtual transitions through higher atomic levels account for a major part of the deviations in the resonator-mode transition frequency caused by simple two-level truncation.

In the L\"owdin full approximation shown in Fig.~\ref{fig4}(c), large errors appear as the coupling scale increases, and the regions in which these errors occur vary nonmonotonically with $\gamma$.
This result shows that rediagonalization of the energy-independent second-order effective Hamiltonian employed here does not necessarily guarantee improved accuracy over the entire parameter range investigated.
The difference between the L\"owdin diagonal and full approximations reflects the distinction between state-specific diagonal corrections and rediagonalization of the second-order matrix including off-diagonal corrections.
However, this result does not imply that off-diagonal physical processes are unnecessary.
In general, when $\epsilon_n$ approaches an eigenvalue of $\hat H_{QQ}$, the energy denominators in Eq.~\eqref{eq:Sigma} can become small, potentially leading to large self-energy contributions.
Furthermore, state mixing caused by rediagonalization of the correction matrix and the range of $P$-space eigenstates retained in the calculation may also affect the accuracy of the approximation.
These contributions cannot be uniquely separated based on Fig.~\ref{fig4} alone.
Therefore, the results presented here demonstrate the limitations of the second-order approximation employed in this study.

Interestingly, Fig.~\ref{fig4}(a) also shows parameter regions in which the relative error of the two-level truncation for the resonator mode decreases as $\gamma$ increases.
However, varying $\gamma$ does not simply change a single measure of anharmonicity; it simultaneously modifies the energy-level separations, wave functions, and coupling matrix elements of the artificial atom.
Moreover, corrections to transition frequencies depend on the difference between the energy corrections of the initial and final states.
Therefore, this tendency cannot be attributed solely to changes in the shunt capacitance or anharmonicity, and the multilevel structure, including the coupling to higher atomic levels, must be taken into account.

Previous studies have discussed the classification of spectral regimes within the Rabi model, including the regime where the Jaynes--Cummings model is valid, the regime where counter-rotating terms can be treated perturbatively, and regimes beyond these approximations \cite{PhysRevA.81.042311,PhysRevA.96.013849}.
In the present study, I have evaluated the accuracy of truncating a multilevel artificial atom to two levels and perturbatively incorporating the contributions of the eliminated atomic levels, thereby identifying a nonperturbative regime associated with the truncation of multilevel atomic degrees of freedom.
These results are also consistent with previous work showing that incorporating the contributions of higher atomic levels into a low-energy effective Hamiltonian can improve the accuracy of the two-level approximation \cite{lamberto_renormalization_2025}.

On the other hand, I find that the accuracy of the approximation can differ substantially between diagonal energy corrections applied to individual eigenstates and rediagonalization of a second-order effective Hamiltonian including off-diagonal matrix elements.
This difference suggests that the treatment of energy dependence and state mixing within the low-energy subspace is important when incorporating the contributions of higher atomic levels.

\section{Resonator spectrum over a wide flux range}

So far, I have analyzed the effects of truncating higher atomic levels on the low-energy spectrum near half a flux quantum.
On the other hand, the resonator spectrum shown in Fig.~\ref{fig1}(c) varies over a wide range of magnetic flux biases.
Except in the immediate vicinity of $f=0.5$, this response predominantly corresponds to the resonator mode and reflects the magnetic-flux dependence of the effective inductance of the artificial atom, which shares a superconducting loop with the resonator.

In principle, the blocks of the circuit Hamiltonian can be constructed at each magnetic flux bias, and the corrections obtained from L\"owdin partitioning can be evaluated in the same manner as in the previous section.
However, over a wide range of magnetic flux biases, the shape of the potential changes substantially from a double-well to a single-well structure.
With the expansion in a harmonic-oscillator or plane-wave basis employed in this study, a large Hilbert space is required to achieve sufficient convergence of the eigenvalues over the entire range, resulting in increased computational cost \cite{Tomonaga_2025}.

I therefore introduce a phenomenological frequency correction into a semiclassical model based on Josephson inductance.
This approach provides an efficient description of the experimental spectrum over a wide magnetic flux range and serves as a means of estimating the initial parameters for fitting the multilevel circuit Hamiltonian.
Furthermore, independently of the role of the additional phenomenological coefficients, examining whether the spectra in different magnetic flux regions can be described using a common set of circuit parameters provides supplementary information for assessing the consistency of the parameters.

Except in the immediate vicinity of $f=0.5$, when the probe angular frequency $\omega_{\mathrm p}$ is sufficiently lower than the excitation frequencies of the artificial atom, I can consider a response in which atomic excitations are suppressed.
In the following, I further adopt a semiclassical approximation in which the phases across the Josephson junctions are treated near the minima of the potential, and describe the response of the artificial atom in terms of Josephson inductance \cite{harris_compound_2009,miyanaga_ultrastrong_2021}.

Defining $L_{\mathrm J}=(\Phi_0/2\pi)^2/E_{\mathrm J}$, the inductance of junction $a$ is given by $L_{\mathrm J}/\cos\varphi_a$.
The effective inductance seen across the $\beta$ junction is evaluated using the following expression:
\begin{equation}
L_{\mathrm A}
=
\frac{L_{\mathrm J}}{
\displaystyle
\beta\cos\varphi_\beta
+\frac{2\cos\varphi_a+\alpha\cos\varphi_\alpha}
{\cos\varphi_a\cos\varphi_\alpha}
}.
\label{eq:LA}
\end{equation}

To provide a physical motivation for the spectral corrections arising from higher atomic levels, I expand Eq.~\eqref{eq:lowdin_second_order} in terms of the eigenstates of the $Q$ space:
\begin{equation}
\Delta\epsilon_n^{(2)}
=
-\sum_k
\frac{
\left|\mel{v_k}{\hat H_{QP}}{u_n}\right|^2
}{
E_k^{(Q)}-\epsilon_n
}.
\label{p}
\end{equation}
The denominator represents the energy difference between the state of interest in the $P$ space and an intermediate state in the $Q$ space.
When $E_k^{(Q)}>\epsilon_n$, this coupling produces a negative energy correction due to level repulsion.
This is consistent with the concept of virtual transitions in the dispersive regime \cite{blais_cavity_2004}.

In this section, rather than rigorously deriving the following correction function from Eq.~\eqref{p}, I introduce a phenomenological function to describe the downward shift and magnetic-flux dependence of the measured resonator spectrum:
\begin{equation}
\begin{aligned}
\delta\omega(f)&=\sum_{i=0}^{1}\delta\omega_i(f),\\
\frac{\delta\omega_i(f)}{2\pi}
&=-\sqrt{A_i^2+
\frac{B_i^2}{1+C_i^2(f-0.5)^2}}.
\end{aligned}
\label{phenom_shift}
\end{equation}
Here, $A_i$ and $B_i$ have dimensions of ordinary frequency, whereas $C_i$ is dimensionless.
The corrected resonator angular frequency is defined as
\begin{equation}
\omega_{\mathrm{res}}^{(\mathrm{ph})}(f)
=\frac{1}{\sqrt{L_{\mathrm A}(f)C_{\mathrm r}}}
+\delta\omega(f).
\label{resonator}
\end{equation}

The coefficients obtained by fitting Eq.~\eqref{phenom_shift} are
$A_0=17.94$~MHz,
$A_1=0.63$~MHz,
$B_0=48.51$~MHz,
$B_1=12.79$~MHz,
$C_0=6.04$,
and $C_1=52.59$.
Since $C_1>C_0$, the contribution with $i=1$ exhibits a magnetic-flux dependence that is more localized near half a flux quantum than that with $i=0$.

The white dotted line in Fig.~\ref{fig1}(c) shows the fitting result including the phenomenological correction, whereas the black dotted line shows the semiclassical result without this correction.
Figure~\ref{fig1}(d) shows the two correction terms, one of which exhibits a strong magnetic-flux dependence near $f=0.5$, while the other exhibits a weaker magnetic-flux dependence over the entire range.
The circuit parameters other than the phenomenological correction coefficients are the same as those used in the analysis of Fig.~\ref{fig1}(e).

\section{Conclusion}

In this study, I have analyzed the spectra of a multilevel superconducting artificial atom coupled to a resonator in the ultrastrong coupling regime, both experimentally and theoretically, and investigated the effects of truncating higher atomic levels on the low-energy spectrum.
Using a multilevel circuit Hamiltonian consistent with the measured spectrum as a reference, I compared simple two-level truncation with corrections based on L\"owdin partitioning.
I showed that energy shifts arising from virtual transitions through higher atomic levels are an important source of the spectral deviations between the two-level model and the multilevel circuit model.

Furthermore, I systematically varied the coupling strength and circuit parameters and evaluated the relative error in $\omega_{02}$, corresponding to the resonator mode.
Even in regions where simple two-level truncation produces large errors, diagonal energy corrections based on L\"owdin partitioning substantially reduce the errors over a wide parameter range.
On the other hand, rediagonalization of the second-order effective Hamiltonian including off-diagonal matrix elements does not necessarily improve the accuracy.
This result indicates the range of applicability of the second-order approximation employed here, rather than a limitation of exact L\"owdin partitioning.
The results also suggest that the accuracy of these approximations cannot be determined solely from the atomic level separations or anharmonicity, but also depends on the coupling matrix elements to higher atomic levels and the differences between the energy corrections of the states involved in the transitions.

When the effective parameters of the Rabi model are refitted, the spectrum of the multilevel circuit Hamiltonian can be reproduced with good accuracy under certain conditions, whereas systematic residual deviations remain for parameters corresponding to the measured circuit.
This indicates not only that the bare circuit parameters and effective Rabi parameters cannot be directly identified with each other, but also that some multilevel effects cannot be absorbed solely by readjusting constant parameters.
Furthermore, the experimental resonator spectrum over a wide range of magnetic flux biases was reproduced by combining a semiclassical model based on Josephson inductance with a phenomenological frequency correction.

These results demonstrate the importance of separately evaluating the accuracy of two-level truncation, the perturbative contributions of virtual transitions through higher atomic levels, and refitting to an effective Rabi model in the quantitative modeling of ultrastrongly coupled quantum circuits.
These comparisons provide guidance for understanding the conditions under which low-dimensional effective models are valid and their limitations in ultrastrongly coupled systems containing multilevel artificial atoms.

\,

\section*{Acknowledgments}

I thank H. Mukai, R. Stassi, S. Masuda, K. Mizuno, and T. Aoki for helpful discussions and comments.
This work was supported by JSPS KAKENHI (Grant Nos. JP22K21294 and JP23K13048),
project JPNP16007, and the MATSUO FOUNDATION.

\bibliography{refs.bib}
\end{document}